\documentclass[journal,onecolumn]{IEEEtran} 

\IEEEoverridecommandlockouts                              

\usepackage{subcaption}
\usepackage{url}
\usepackage{graphicx} 

\usepackage{mathptmx} 

\usepackage{times} 

\usepackage{amsmath} 
\usepackage{amssymb}  
\usepackage{bm}
\usepackage{algorithm}
\usepackage{algpseudocode}
\usepackage{multirow}

\title{\LARGE \bf
An optimal reeling control strategy for pumping airborne wind energy systems without wind speed feedback
}

\author{Andrea Berra and Lorenzo Fagiano 
\thanks{The authors are with the Dipartimento di Elettronica, Informazione e Bioingegneria, Politecnico di Milano, Milano, Italy. E-mail addresses:
	{\tt\small lorenzo.fagiano@polimi.it, andrea.berra@mail.polimi.it }}}

\begin{document}

\maketitle
\thispagestyle{empty}
\pagestyle{empty}

\begin{abstract}
Pumping airborne wind energy (AWE) systems employ a kite to convert wind energy into electricity, through a cyclic reeling motion of the tether. The problem of computing the optimal reeling speed for the sake of maximizing the average cycle power is considered. The difficulty stems from two aspects: 1) the uncertain, time- (and space-) varying nature of wind speed, which can not be measured accurately, and 2) the need to consider, in the same optimization problem, the different operational phases of the power cycle. 
A new, model-based approach that solves this problem is proposed. In the design phase, a model of the AWE system is employed to collect data pertaining to the cycle power obtained with various reel-in/reel-out speed pairs, assuming known wind speed. Then, a nonlinear map, identified from these data, is used as cost function in an optimization program that computes the best reel-in and -out speed pairs for each wind speed. Finally, the optimization results are exploited to infer the link between optimal reeling speed and tether force, which are both measured with high accuracy. Such a link is used to design a feedback controller that computes the reeling speed based on the measured tether force, in order to converge on the optimal force-speed manifold. Simulation results with a realistic model illustrate the effectiveness of the approach.
\end{abstract}

\section{Introduction}
Airborne Wind Energy (AWE) generators are autonomous systems that employ a kite, tethered to a ground station, to extract kinetic energy from wind \cite{FaMi12,Schmehl2018}. In pumping AWE, the kite carries out a power generation cycle composed of two phases: a \textit{traction} one, where the tether is reeled-out from a ground winch at low speed and under high force, and a \textit{retraction} one, where the tether is reeled-in at higher speed and under much lower force. The reeling motion is managed by the ground station, with a feedback controller acting on the electric machine coupled with the winch. At the same time, a flight controller is in charge of carrying out suitable trajectories in the two phases, as well as in transitions between them. The flight trajectories differ significantly among the operational phases, since fast crosswind motion is required during traction, while a non-crosswind motion is more effective during retraction. The \textit{cycle power} is the average power generated in a single pumping cycle, while \textit{traction power} and \textit{retraction power} refer to the average values obtained during each of the two corresponding phases.\\
To obtain the highest conversion efficiency, cycle power shall be maximized during operation (subject to constraints such as maximum tether force and tether speed limits), however this is not trivial to obtain. The first reason is that the optimal reeling speed depends on the flight path of the kite, and in particular on the wind speed encountered by the latter, which is generally time- and space-varying and is not accurately measured. Thus, wind speed can not be used as feedback variable by the reeling controller: strategies that compute the optimal pumping cycle (see e.g. \cite{Houska2010,Fagiano2011a}) assuming known wind speed  can not be directly translated into a practical feedback controller. To overcome this problem, reeling control techniques with different feedback variables have been proposed, and some have been tested experimentally, see e.g. \cite{Erhard2015,Zgraggen2016,WOOD2018981}).
In general, these approaches exploit a model of the system to devise a link among measurable quantities (e.g., tether force, tether speed, or kite airspeed) during optimal operation with known wind speed, which can be easily assessed analytically or in simulation. Such a link is then used to devise rather simple controllers that make the system operate on the found optimal manifold. In \cite{Erhard2015}, the reel-out speed is computed as a function of the measured airspeed in order to achieve maximum traction power, while the reel-in speed is computed as a function of the angle between the tether and an inertial reference axis. In \cite{Zgraggen2016}, the generator torque during reel-out is computed as a function of the reel-out speed to maximize traction power. Albeit very effective, these solutions are suboptimal with respect to cycle power maximization: this is due to the presence of the retraction phase, which is dealt with separately. Indeed, the need to consider the different operational phases altogether is the second main reason that contributes to the problem difficulty.\\
In this paper, we propose a new, systematic approach to design feedback reeling controllers for both phases, that starts from the same general idea as \cite{Erhard2015,Zgraggen2016} but overcomes the mentioned problems. We employ a model of the system to estimate the response surface of power cycle as a function of reel-in and -out speeds, for different wind speed values. Then, using such a response surface we compute the manifold of optimal reeling speeds and corresponding optimal traction force values as parametrized by the wind speed. Finally, we derive a feedback law where the reference reeling speed is computed based on the measured tether force, such that these two variables converge to the found manifold. Therefore, the resulting control strategy employs tether force and speed as feedback variables, which are readily available. Without loss of generality, we consider a soft kite system with a single tether, such as those of \cite{Baayen2012,Erhard2015}, but the general methodology can be used also for systems with flexible wings and more tethers, such as \cite{Bormann2014,kitenergy,Zgraggen2016}. 
Simulation tests with a widely used AWE model show that the proposed approach achieves optimal performance, i.e. the same that would be obtained with an optimal reeling speed allocation assuming exact knowledge of the wind speed.

\section{System model, pumping cycle control, and problem formulation}\label{S:model,control,problem}

\begin{figure}[htbp]
	\centering
	\includegraphics[width=\columnwidth]{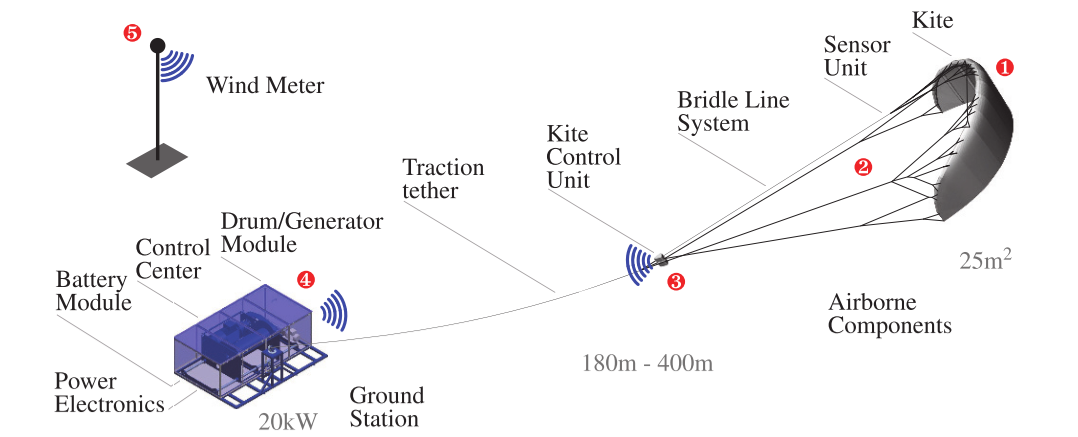}
	\caption{Layout and main components of the considered pumping AWE system. Source: \cite{delft_real_system}.}
	\label{fig:delft_real_system}
\end{figure}

We consider an AWE system with a flexible wing, an airborne kite steering unit (KSU), and a single tether that connects them to a ground station, equipped with a winch linked to an electric machine, like those developed by the companies Skysails \cite{skysails,Erhard2015} and Kitepower \cite{kitepower} and the research group at TU Delft \cite{Schmehl2013}, see Fig. \ref{fig:delft_real_system} for an overview. 
	

\subsection{System model} \label{SS:system model}
Let us denote with $\tau\in\mathbb{R}$ the continuous time variable. We consider a well-established model of the kite, see, e.g., \cite{CaFM09c} and references therein, adopting spherical coordinates: the kite's position is expressed by its elevation angle $\vartheta(\tau)$, azimuth angle $\varphi(\tau)$, and distance from the ground station, which is the origin of an inertial coordinate system. We connect the kite model to a new multi-body model of the tether, derived as an extension of the one described in \cite{BologniniFagiano2020} and introduced next.  The tether is  modeled as a chain of $N_t$ inner nodes, each one with mass $m_{t}(\tau)$ computed as:
\begin{equation}\label{eq:tether_parameters}
\begin{array}{rcl}
m_{t}(\tau) & = & \dfrac{L_t(\tau)\rho_t}{N_t}, 
\end{array}
\end{equation}
where $\rho_t$ is the tether mass per unit of length and $L_t(\tau)$ is the nominal tether length, computed as
\[
L_t(\tau)=r_w\theta_w(\tau),
\]
with $r_w$ being the drum's radius and $\theta_w(\tau)$ the winch angular position (a single-layer of tether is assumed to be coiled around the drum, and without loss of generality we assume that $\theta_w(\tau)=0$ corresponds to the tether fully coiled on the drum). Each node has associated three-dimensional position and velocity vectors, $\bm{p}_{t,l}(\tau), \dot{\bm{p}}_{t,l}(\tau)\in\mathbb{R}^3,l=1,\ldots,N_t$, where $\tau\in\mathbb{R}$ is the continuous time variable. By adding the points at the extremes, assumed to be fixed to the kite and to the ground station,  we thus have $N_t+1$ segments, and each inner node is subject to its own weight, to the aerodynamic force, and to the forces applied by the two neighboring tether segments. The aerodynamic force is modeled by a vector $\boldsymbol{F}_{a,t,l}(\tau)$ computed as:
\begin{equation}\label{eq:tether_aero}
\begin{array}{rcl}
\boldsymbol{F}_{a,t,l}(\tau) & = & \dfrac{1}{2}\rho A_{t,l}^\bot(\tau)\|\boldsymbol{W}(\tau)-\dot{\bm{p}}_{t,l}(\tau)\|(\boldsymbol{W}(\tau)-\dot{\bm{p}}_{t,l}(\tau)), 
\end{array}
\end{equation}
where $\|\cdot\|$ is the 2-norm, $\boldsymbol{W}_l(\tau)$ is the absolute wind speed at the current position of tether segment $l$, $\rho$ is the air density, $A_{t,l}^\bot(\tau)$ is the projection of the surface of tether segment $l$ on a plane perpendicular to the apparent wind, i.e. to vector $(\boldsymbol{W}_l(\tau)-\dot{\bm{p}}_{t,l}(\tau))$.
Internal tether forces include elastic and friction ones. In particular,
the tether segments are modeled as nonlinear springs, which can transfer forces only when the tether segment is taut, in parallel with linear dampers. We first compute the nominal length and spring constant of each segment, $\ell_{t,l}(\tau)$ and $k_{t,l}(\tau)$, respectively, as:
\begin{equation}\label{eq:tether_parameters_2}
\begin{array}{rcl}
\ell_{t,i}(\tau) &=& \dfrac{L_i(\tau)}{N_t+1}\\
k_{t,i}(\tau) & = & \dfrac{\overline{F}_t}{\overline{\varepsilon}_t\ell_t(\tau)}
\end{array}
\end{equation}
where  $\overline{F}_t$ is the maximum tether load in the elastic deformation regime and $\overline{\varepsilon}_t$ the corresponding elongation. Considering node $l$, the applied nonlinear elastic forces read (time dependence of the involved variables is omitted for notational simplicity):\small
\begin{equation}\label{eq:node_i_elastic}
\begin{array}{rcl}
\bm{F}_{e,l+1} &=& \max\left(0,k_t\left(\|\bm{p}_{t,l+1}-\bm{p}_{t,l}\|-\ell_t\right)\right)\dfrac{\bm{p}_{t,l+1}-\bm{p}_{t,l}}{\|\bm{p}_{t,l+1}-\bm{p}_{t,l}\|}\\
\bm{F}_{e,l-1} &=& \max\left(0,k_t\left(\|\bm{p}_{t,l-1}-\bm{p}_{t,l}\|-\ell_t\right)\right)\dfrac{\bm{p}_{t,l-1}-\bm{p}_{t,l}}{\|\bm{p}_{t,l-1}-\bm{p}_{t,l}\|},
\end{array}
\end{equation}\normalsize
where the saturation to zero implies that no elastic force is present when the distance between two nodes is smaller than the nominal segment length. The friction forces are supposed to act only in axial direction and are computed as:\small
\begin{equation}\label{eq:node_i_friction}
\begin{array}{rcl}
\bm{F}_{f,l+1} &=& \beta_t\dfrac{\left(\dot{\bm{p}}_{t,l+1}-\dot{\bm{p}}_{t,l}\right)^T\left(\bm{p}_{t,l+1}-\bm{p}_{t,l}\right)}{\|\bm{p}_{t,l+1}-\bm{p}_{t,l}\|^2}\left(\bm{p}_{t,l+1}-\bm{p}_{t,l}\right)\\
\bm{F}_{f,l-1} &=& \beta_t\dfrac{\left(\dot{\bm{p}}_{t,l-1}-\dot{\bm{p}}_{t,l}\right)^T\left(\bm{p}_{t,l-1}-\bm{p}_{t,l}\right)}{\|\bm{p}_{t,l-1}-\bm{p}_{t,l}\|^2}\left(\bm{p}_{t,l-1}-\bm{p}_{t,l}\right)
\end{array}
\end{equation}\normalsize
where $\beta_t$ is the constant friction coefficient of each tether segment. 
For the $l$-th node, the equations of motion thus read:
\begin{equation}\label{eq:node_i_motion}
\ddot{\bm{p}}_{t,l} =\dfrac{\left(\bm{F}_{e,l+1}+\bm{F}_{e,l-1}+\bm{F}_{f,l+1}+\bm{F}_{f,l-1}\right)}{m_t}-\left[
\begin{array}{c}
0\\0\\g
\end{array}\right]
\end{equation}
where $g$ is the gravity acceleration, $\boldsymbol{F}_f$ represent friction forces and $\boldsymbol{F}_e$ elastic ones. For the first and last nodes, i.e., $l=1$ and $l=N_t$, the position and velocity vectors of the tether ends are considered in the computation of tether forces. These are equal to the position and velocity vectors of the devices attached to the tether ends, i.e., those of the kite on one end and of the ground winch on the other end. In turn, the kite and the winch are subject to the elastic and friction force vectors related to the first and last segment, with a minus sign, thus coupling the tether model with those of the connected systems.\\
Finally, the winch is modeled by a rather standard, linear time invariant rotational mechanical system attached to the electric machine, whose torque $T(\tau)$ is assumed to be directly controlled (i.e., current loop dynamics are neglected for the sake of this research):
\begin{equation}\label{eq:node_i_friction}
\begin{array}{rcl}
\ddot{\theta}_w(\tau)&=&\dfrac{-\beta_w\dot{\theta}_w(\tau)+T(\tau)}{J_w}
\end{array}
\end{equation}
where  $\beta_w,\,J_w$ are, respectively, the viscous friction coefficient and moment of inertia of the winch (also including the friction and inertia of the gearbox and of the electric machine). 


\subsection{Pumping cycle control}\label{SS:pumping control}
\begin{figure}[htbp]
	\centering	
	\includegraphics[width=.5\textwidth]{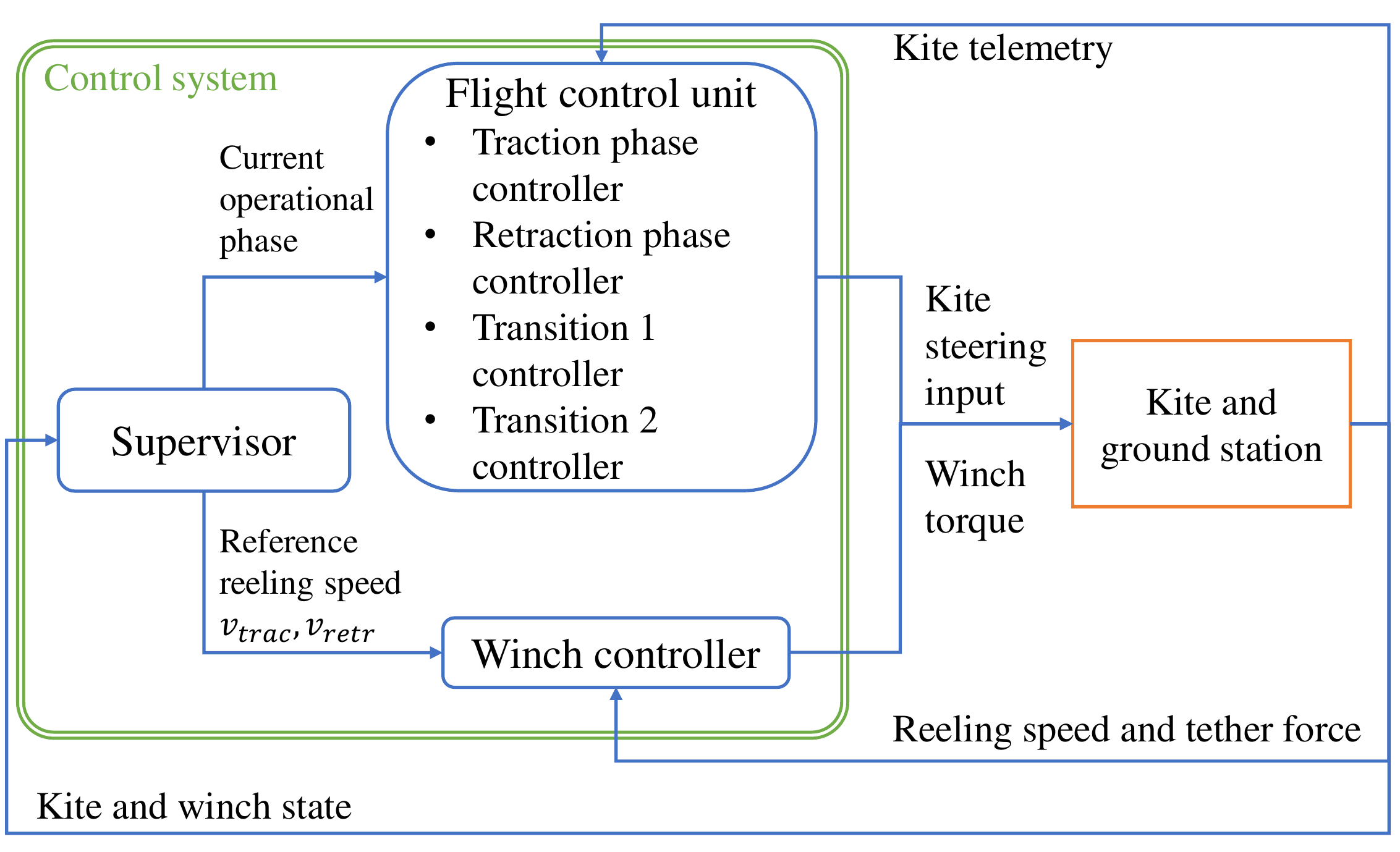}
	\caption{Layout of the considered control structure for pumping AWE systems with soft kite.}
	\label{fig:plant}
\end{figure}

The AWE generator is managed by a hierarchical and distributed control system, presented in Fig. \ref{fig:plant}, featuring two low-level controllers (the flight control unit and the winch control unit) and a supervisor.  Onboard, the KSU steers the kite via an actuator that acts on the bridles, as commanded by the flight control unit. Feedback is provided by an inertial measurement unit and possibly an onboard airspeed sensor. The onboard control system we use in this work exploits velocity angle (or course angle) feedback and a navigation strategy based on target points, as described in \cite{FZMK14,Zgraggen2016} and similar to \cite{Erhard2015}. At the same time, on ground, the winch controller regulates the reeling motion of the tether through the torque exerted by electric machine. The reeling strategy features force, speed, and position feedback loops. Usually, reeling control is achieved either by tracking a torque reference, computed on the basis of the measured reeling speed \cite{Zgraggen2016}, or by tracking a reeling speed reference, computed on the basis of the tether force. Winch position feedback is employed, together with the kite telemetry, by the supervisor, which manages phase transitions and sends information on the current operational phase to the flight and winch controllers, in order to coordinate them. 
In particular, the power cycle includes two main phases:
\begin{enumerate}
	\item \textbf{Traction phase}. The system generates electricity by reeling out the tether from an initial length $\underline{L}_t$, while the kite is steered to carry out figure-of-eight paths to maximize the tether force. 
	\item \textbf{Retraction phase}. When the tether reaches a chosen length $\bar{L}_t>\underline{L}_t$, the winch reels in until $L_t(\tau)\leq\underline{L}_t$, in order to start another production cycle. In this phase, the flight controller steers the kite into a so-called ``low power maneuver'', i.e. the kite is steered to the side of the wind window at relatively large elevation and azimuth angles, see e.g. \cite{FaMi12}, to minimize traction force.
\end{enumerate}
Moreover, two  transition phases are carried out with roughly constant tether length: \textit{Transition 1} aims to steer the kite from the terminal conditions at the end of the traction phase (high speed and large traction force) to suitable starting conditions for the retraction (low speed and , and vice-versa \textit{Transition 2} brings the kite from the terminal conditions of the retraction to the starting conditions of the next traction phase.\\
The high-level control layer switches these phases according to logical conditions on the kite position and speed, as detailed e.g. in \cite{Zgraggen2016,CaFM09c}, see Fig. \ref{fig:plant}. 

\begin{figure}[htbp]
	\centering
	\includegraphics[width=.5\textwidth]{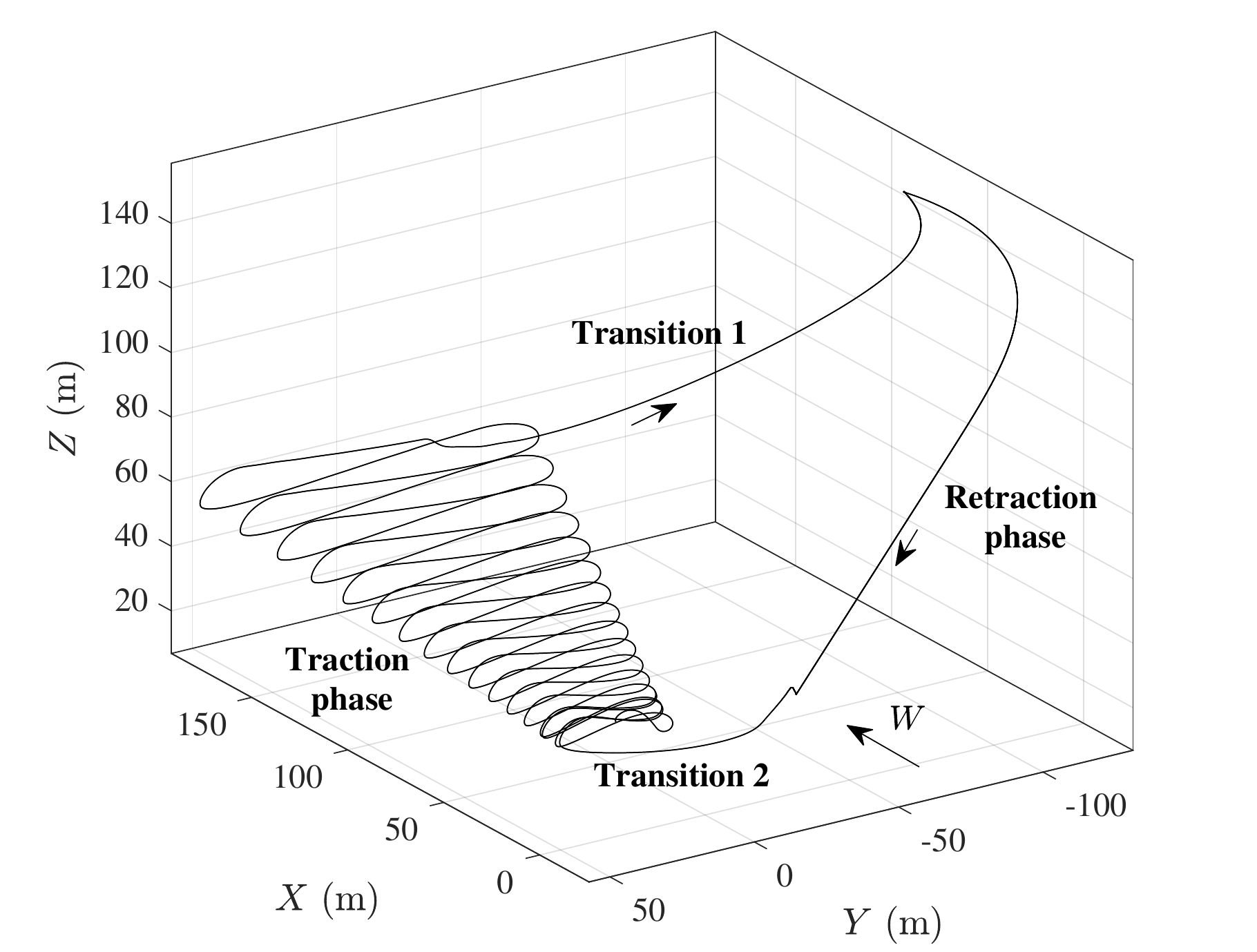}
	\caption{Example of kite path during one pumping cycle obtained with the considered control structure.}
	\label{fig:3dpath}
\end{figure}

With the described control structure, a typical example of kite path obtained in one production cycle is shown in Fig. \ref{fig:3dpath}. 

\subsection{Problem formulation}\label{SS:problem formulation}

This paper focuses on the design of the reeling strategy, in particular the choice of the reference tether speed values that the winch controller shall track in the traction and retraction phases, indicated respectively as $v_{trac}(\tau)$ and $v_{retr}(\tau)$. To introduce the problem, let us consider the cycle power:
\begin{equation}
P_{cycle}=\dfrac{\int\limits_{\tau_s}^{\tau_e}P(\tau)d\tau}{\tau_e-\tau_s},
\label{eq:power}
\end{equation}
where $P(\tau)$ is the winch mechanical power and $\tau_s,\,\tau_e$ the start and end time of the pumping cycle. In this paper we consider mechanical power for simplicity and without loss of generality: our approach can be applied in a straightforward way also considering the electrical power, e.g. by including a model of the conversion losses. Clearly, the cycle power depends on the absolute wind speed $\boldsymbol{W}(\tau)$ and on the kite and tether trajectories. Let us assume that the kite path lies in a-priori-optimized regions on the azimuth-elevation plane, in particular in the so-called ``power zone'' during traction and in a ``parking zone'' during retraction, see Fig. \ref{fig:3dpath}. This assumption is reasonable and corresponds to a well-established approach employed in the literature \cite{JeSc14,Erhard2015,Zgraggen2016}. Then, the most relevant remaining quantities for the sake of cycle power optimization are the wind speed and the reel-in and reel-out speed values. The latter two have the following main effects:
\begin{itemize}
	\item they directly affect the mechanical power with a cubic relationship, since power is given by tether force times speed, and tether force depends quadratically on tether speed (as it is mainly due to aerodynamic forces developed by the kite);
	\item they affect the duty cycle of pumping operation, since the duration of the traction and retraction phases is inversely proportional to the corresponding reeling speeds.
\end{itemize}
For a given absolute wind speed magnitude, indicated as $v_w=\|\boldsymbol{W}\|$, one can resort to simplified equations or to dynamical models to compute the corresponding optimal values of $v_{trac}(v_w), v_{retr}(v_w)$ to maximize the cycle power, \cite{Houska2010,Fagiano2011}. This approach is useful to derive an upper bound on the performance that the system can achieve, however it can not be used directly to design the reeling strategy, because in practice the wind speed at the kite position is not accurately measured, and a small measurement/estimation error can produce a rather large decrease of generated power, due to the cubic power-speed dependence.

Moreover, an important aspect that can not be considered with simplified equations is whether the chosen reeling speeds $v_{trac}, v_{retr}$ return a feasible periodic system behavior or not. Here, feasibility is intended as the satisfaction of all physical operation limits, such as the kite's position being always in user-defined flight zones, tether force being below a safety threshold, etc. In order to include this aspect in our problem, let us indicate with a binary variable $s$ whether the model simulation returns a repetitive pumping cycle without any constraint violation ($s=1$) or not ($s=0$). In the considered setup, the value of $s$ depends only on the wind speed $v_w$ and the reeling speeds $v_{trac},\,v_{retr}$. Thus, for a given wind speed value we can define the set $V_{reel}(v_w)$ of admissible reeling speed pairs as:
\begin{equation}\label{eq:admissible speeds set}
V_{reel}(v_w)=\left\{(v_{trac},v_{retr}):s\left(v_w,v_{trac},v_{retr}\right)=1\right\}.
\end{equation}
To characterize this set analytically is challenging, since the constraints are evaluated through the dynamical simulation of a rather complex nonlinear system controlled by the hierarchical approach described in Section \ref{SS:pumping control}. However, our approach is based on sampling the search space $(v_{trac}, v_{retr})$, so that the explicit computation of $V_{reel}(v_w)$ is not required. Moreover, as a matter of fact, one can resort to application-related expertise to establish sensible upper and lower bounds on $v_{trac}$ and $v_{retr}$ (e.g., up to a certain fraction of $v_w$ and within the speed limits of the electric machine), such that the resulting hyper-rectangle is contained in $V_{reel}(v_w)$.

Therefore, the problem we address here is that of deriving a reeling control strategy that computes the reference speeds $(v_{trac}(\tau), v_{retr}(\tau))\in V_{reel}(v_w(\tau))$, without relying on a measurement of the wind speed magnitude, such that the resulting cycle power is maximized. The approach we propose to solve this problem is described next.

\section{Proposed reeling control design method}\label{S:control approach}

\subsection{Design method}\label{SS:design method}
The key idea of our approach is to first compute, via numerical optimization applied to the full pumping system model, the optimal operating conditions assuming known wind speed $v_w$. Then, we  derive the manifold, in the space of measurable quantities, that contains the optimal feasible working points corresponding to $v_w$ values in an interval of interest $V_w=[\underline{v}_w,\,\bar{v}_w]$. In particular, we consider tether speed and force as done already in previous studies, where however only the traction power was maximized relying on simple static equations \cite{Erhard2015,Zgraggen2016}.
As we show next, the optimal force-speed pairs belonging to the manifold correspond to unique maxima of cycle power for each wind speed in the considered interval. Finally, we derive a feedback strategy that sets the reference reeling speed based on tether force feedback, such that these two quantities converge to the obtained manifold, thus guaranteeing optimal operation without the need of wind speed measurement. Conversely, one could infer the wind speed from the close loop force-speed behavior, even though wind speed estimation is not the focus of this paper.

More precisely, our design method features the following steps.\\

\noindent\textbf{Off-line phase}.
\begin{enumerate}
	\item Select a finite number of values $\tilde{v}_w^{(i)}\in V_w,\,i=1,\ldots,N$ (e.g., by uniform gridding). For each $\tilde{v}_w^{(i)}$, create a data-set of cycle power values by simulating full pumping cycles for a finite number $M$ of reference reeling speed pairs $(\tilde{v}_{trac},\tilde{v}_{retr})^{(j)},\,j=1,\ldots, M$, chosen again by gridding over physically sensible intervals.  
	Let us denote with $\tilde{P}^{(i,j)}$ the cycle power \eqref{eq:power} obtained with wind speed $\tilde{v}_w^{(i)}$ and reference reeling speeds $(\tilde{v}_{trac},\tilde{v}_{retr})^{(j)}$. If a chosen pair $(\tilde{v}_{trac},\tilde{v}_{retr})^{(j)}$ falls outside the set $V_{reel}(\tilde{v}_w^{(i)})$ (see \eqref{eq:admissible speeds set}), which can be easily checked based on the simulation results, assign to such a pair a very low value of $\tilde{P}^{(i,j)}\ll0$;
	\item For each $i=1,\ldots,N$, carry out the following tasks:
	\begin{enumerate}
		\item Use the data $\tilde{P}^{(i,j)},\,(\tilde{v}_{trac},\tilde{v}_{retr})^{(j)},\,j=1,\ldots, M$ to derive a response surface $f^{(i)}:\mathbb{R}^2\rightarrow\mathbb{R}$ that provides an estimate of the power cycle $P$ obtained with wind speed $v_w^{(i)}$ as a function of the reel-out and reel-in reference speeds. This can be done by choosing a suitable parametric model $\hat{P}=f^{(i)}((v_{trac},v_{retr}),\boldsymbol{\theta})$ with parameters $\boldsymbol{\theta}\in\mathbb{R}^{n_{\boldsymbol{\theta}}}$ and solving the following optimization problem:
		\begin{equation}\label{eq:fitting problem}
		\min\limits_{\boldsymbol{\theta}}\sum\limits_{j=1}^M\left(\tilde{P}^{(i,j)}-f^{(i)}\left((\tilde{v}_{trac},\tilde{v}_{retr})^{(j)},\boldsymbol{\theta}\right)\right).
		\end{equation}
		In this work, we chose $f$ as a linear combination of Gaussian basis functions centered at each data point, but we obtained equivalently good results with polynomials and neural networks. A regularization term can be also included to limit over-fitting. Let us denote with $\boldsymbol{\theta}^{(*,i)}$ the obtained minimizer to \eqref{eq:fitting problem}. Note that the response surface will take into account unfeasible speed pairs, since these correspond to very low power values, see point 1).
		\item Use the derived power cycle model $f^{(i)}\left((v_{trac},v_{retr}),\boldsymbol{\theta}^{(*,i)}\right)$ to compute the optimal reference reeling speed values as:
		\begin{equation}\label{eq:optimization problem}
		\begin{array}{c}
		\max\limits_{(v_{trac}, v_{retr})}f^{(i)}\left((v_{trac},v_{retr}),\boldsymbol{\theta}^{(*,i)}\right)
		\end{array}
		\end{equation}
		Let us denote with $\left(v_{trac}, v_{retr})^{(*,i)}\right)$ the obtained minimizer to \eqref{eq:optimization problem}. 
		\item Run a further simulation to evaluate  whether $\left(v_{trac}, v_{retr})^{(*,i)}\right)\in V_{reel}(\tilde{v}_w^{(i)})$. If not, add this pair to the data-set with a corresponding power value $\ll 0$ and go to a).		
	\end{enumerate}
	\item Consider now the speed pairs  $\left(v_{trac}, v_{retr}\right)^{(*,i)},\,i=1,\ldots,N$. From the corresponding simulations, collect the average tether force values in the traction and retraction phases, denoted as $(F_{trac}, F_{retr})^{(*,i)},\,i=1,\ldots,N$:
	\begin{equation}\label{eq:average forces}
	\begin{array}{rcl}
	F_{trac}^{(*,i)}&=&\left(\int\limits_{\tau_{trac,s}}^{\tau_{trac,e}}F^{(*,i)}(\tau)d\tau\right)/\left(\tau_{trac,s}-\tau_{trac,e}\right)\\
	F_{retr}^{(*,i)}&=&\left(\int\limits_{\tau_{retr,s}}^{\tau_{retr,e}}F^{(*,i)}(\tau)d\tau\right)/\left(\tau_{retr,s}-\tau_{retr,e}\right)\\
	\end{array}	
	\end{equation}
	where $F^{(*,i)}(\tau)$ is the simulated course of the tether force with wind speed $\tilde{v}_w^{(i)}$ and reference reeling speeds $\left(v_{trac}, v_{retr}\right)^{(*,i)}$,  $\tau_{trac,s},\,\tau_{trac,e}$ are the start and end time of the traction phase, and $\tau_{retr,s},\,\tau_{retr,e}$ those of the retraction one;
	\item Using the data $\left(v_{trac}^{(*,i)},\,F_{trac}^{(*,i)}\right)$ and $\left(v_{retr}^{(*,i)},\,F_{retr}^{(*,i)}\right),\,i=1,\ldots,N$, fit the following models via linear least squares:
	\begin{equation}\label{eq:optimal manifold}
	\begin{array}{rcl}
	F^*_{trac}&=&K_{trac}v_{trac}^{*2}\\
	F^*_{retr}&=&K_{retr}v_{retr}^{*2}\\
	\end{array}	
\end{equation}	
	where $K_{trac},\,K_{retr}$ are coefficients to be identified, that link the tether force to the square of the reeling speed.\\
\end{enumerate}

\noindent\textbf{On-line phase}.\\
Implement Algorithm \ref{opt_vel_algo} to compute the reference speed given to the winch control system, where we assume that a suitable measurement and control sampling frequency is adopted and denote with $k\in\mathbb{Z}$ the discrete time variable. The only tuning parameter in the algorithm is the scalar $\lambda>0$, see the discussion below for tuning guidelines.\\
\begin{algorithm}[hbt!]
	\caption{Reference velocity computation}
	\begin{algorithmic}[1]
		\State{At each sampling time $k$ measure the tether force $F(k)$ and speed $v(k)=r_w\dot{\theta}_w(k)$}
		\If{Traction phase}
		\State{Compute \(F^*(k)=K_{trac}v(k)\)}
		\If {\(F(k)>F^*(k)\)}
			\State{\(v_{trac}(k+1)=v_{trac}(k)+\lambda\)}
		\ElsIf {\(F(k)<F^*(k)\)}
			\State{\(v_{trac}(k+1)=v_{trac}(k)-\lambda\)}
		\Else
			\State{\(v_{trac}(k+1)=v_{trac}(k)\)}
		\EndIf
		\ElsIf{Retraction phase}
		\State{Compute \(F^*(k)=K_{retr}v(k)\)}
		\If {\(F(k)>F^*(k)\)}
		\State{\(v_{retr}(k+1)=v_{retr}(k)+\lambda\)}
		\ElsIf {\(F(k)<F^*(k)\)}
		\State{\(v_{retr}(k+1)=v_{retr}(k)-\lambda\)}
		\Else
		\State{\(v_{retr}(k+1)=v_{retr}(k)\)}
		\EndIf		
		\EndIf
	\end{algorithmic} 
	\label{opt_vel_algo}
\end{algorithm}
\begin{figure}[!htb]
\centering
	\begin{subfigure}{.7\columnwidth}
		\centering
		\includegraphics[width=\columnwidth]{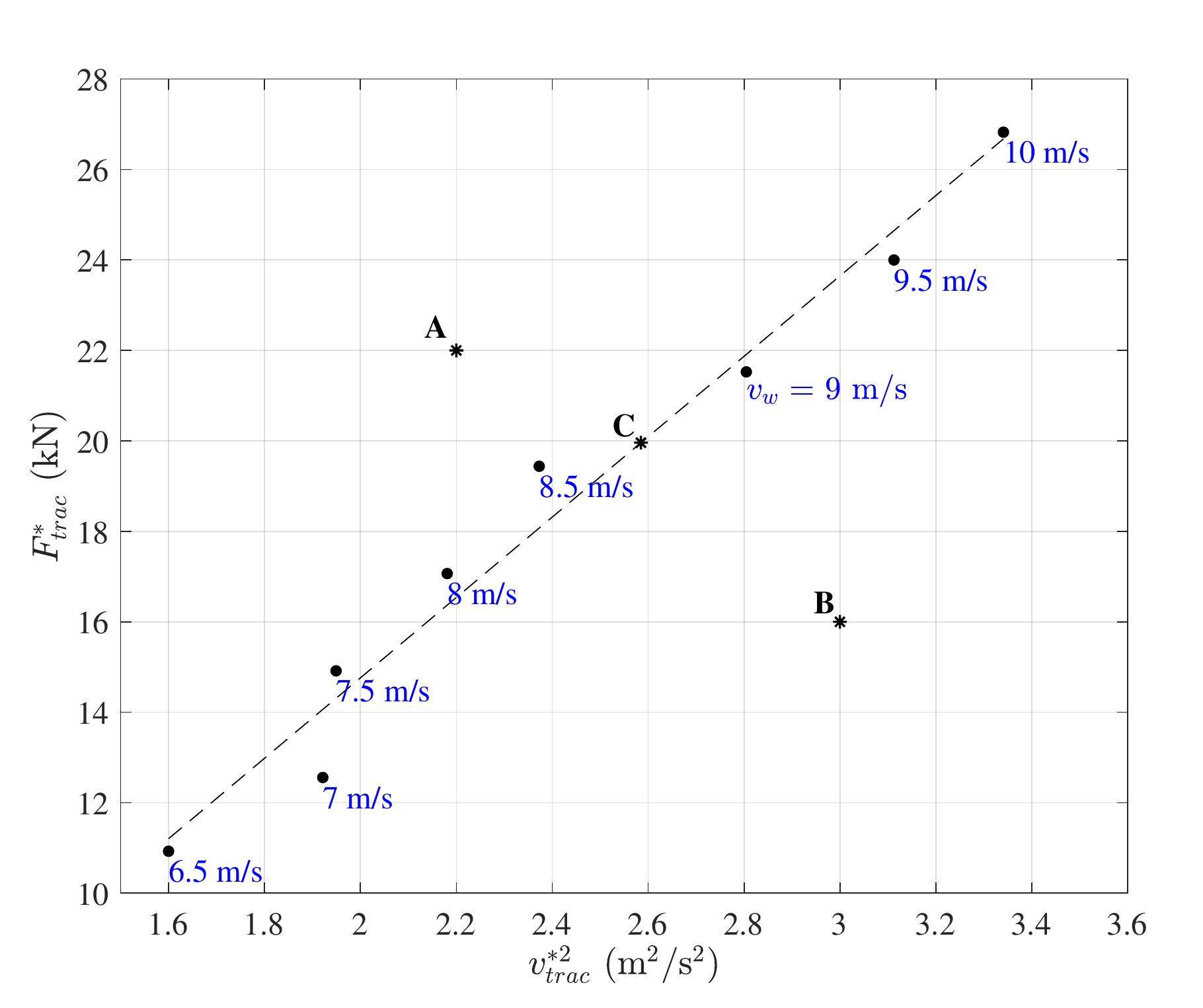}
	\end{subfigure}\\
	\begin{subfigure}{.7\columnwidth}
		\centering
		\includegraphics[width=\columnwidth]{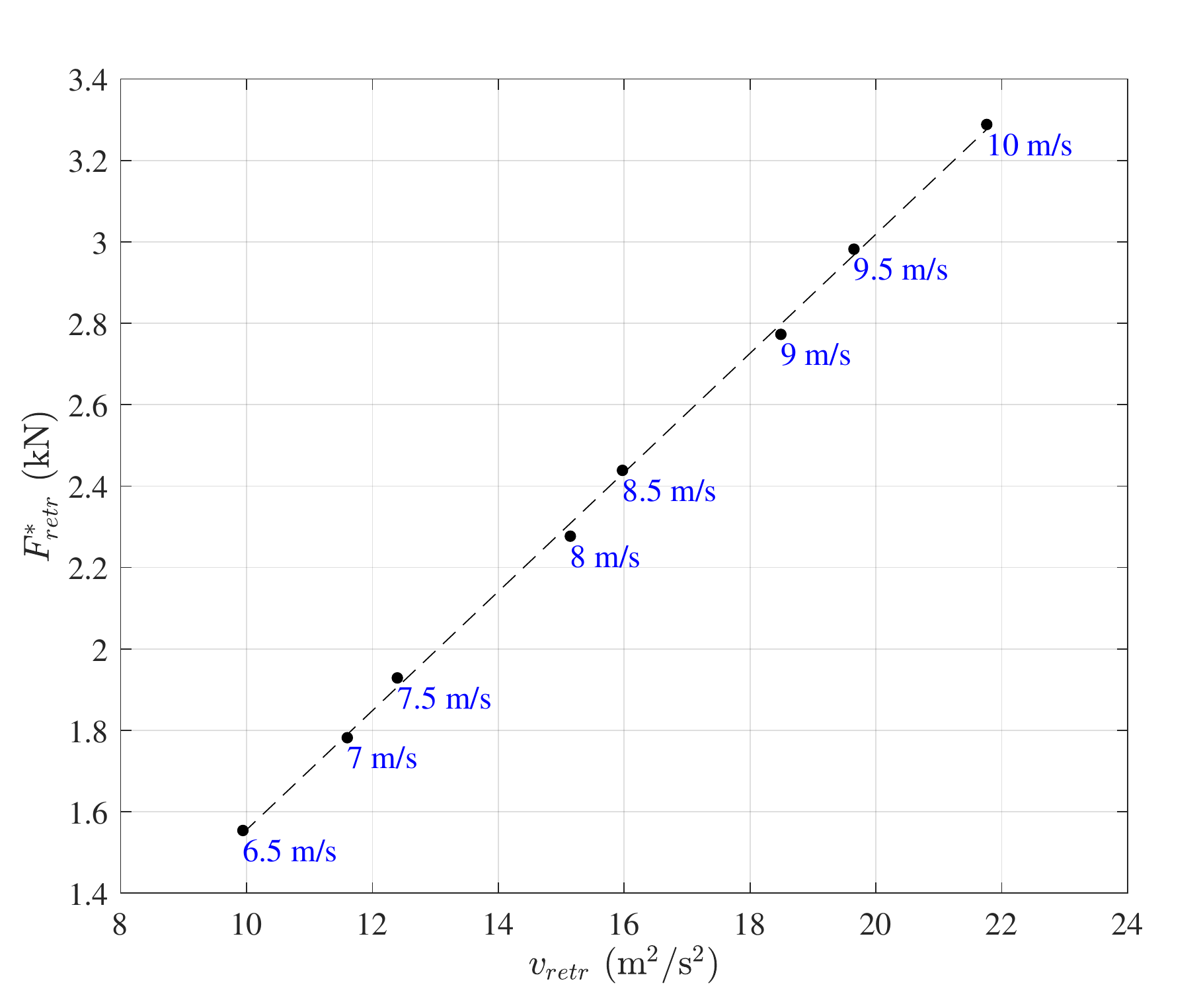}
	\end{subfigure}
	\caption{Simulation results: data points  obtained by solving \eqref{eq:optimization problem} for different \(v_{w}\) values (black dots) and obtained optimal manifolds \eqref{eq:optimal manifold} of squared reference speed and tether force (dashed lines). The wind speed value corresponding to each data point is indicated in blue. Upper plot: traction phase; lower plot: retraction phase. In the upper plot, the three points A, B, and C correspond to the three possible cases in Algorithm \ref{opt_vel_algo}.\label{fig:optimal_F_v2}}	
\end{figure}
\subsection{Discussion}\label{SS:procedure discussion}
Algorithm \ref{opt_vel_algo} is a feedback strategy where the reference reeling speed is gradually adjusted, by discrete steps equal to $\lambda$, in order to converge to a neighborhood of the optimal manifold $F^*_{trac}(v_{trac})$ (or $F^*_{retr}(v_{retr})$ for the retraction phase) derived off-line, see \eqref{eq:optimal manifold}. A small $\lambda$ value leads to a very smooth but possibly slow convergence to the optimal manifold, while an excessively large value may lead to instability and chattering. To improve the behavior of Algorithm 1, one can also resort to a change of reeling speed that is proportional to the difference between the actual force and the optimal one. To understand the rationale behind Algorithm \ref{opt_vel_algo}, it is useful to first analyze the manifolds of optimal force-squared speed pairs obtained by solving \eqref{eq:optimization problem} for different wind speeds, presented in Fig. \ref{fig:optimal_F_v2}. Consider for example the traction phase. It can be noted that the data points are well-aligned along lines in the $(v_{trac}^{*2},F_{trac}^*)$ plane. This is consistent with existing results on traction phase power maximization, however with smaller gain here, since the optimal reel-out speed for the sake of cycle power optimization turns out to be much smaller than the value $v_w/3$ known and employed in the literature \cite{Erhard2015,Zgraggen2016} to maximize traction power. Moreover, we note that a similar behavior is observed for the retraction phase: such a linear link between squared optimal reel-in speed $v_{retr}^{*2}$ and optimal average tether force $F_{retr}^*$ can be expected from first principles also in this case, however the precise gain for cycle power maximization is not easily found a priori. On the other hand, our simulation-based approach allows one to compute both coefficients $K_{trac},\,K_{retr}$ systematically and considering the full pumping system model. We expect a similar qualitative behavior to hold also with more sophisticated models, that can be directly used in our procedure.\\
Now, we illustrate the functioning of Algorithm \ref{opt_vel_algo} considering three different scenarios based on the difference between the optimal force, \(F^*(k)\), and the measured one, \(F(k)\). Again, we take the traction phase as example, the same considerations hold for the retraction one. In the first scenario (point A in Fig. \ref{fig:optimal_F_v2}, upper plot) the measured force  is larger than the optimal one, and the system is not working on the optimal manifold. In this case, considering that a larger speed leads to a reduction of tether force, the Algorithm commands to increase the reel-out speed (line 5 of Algorithm \ref{fig:optimal_F_v2}): as a consequence, the operating point will move towards larger speed and lower force, hence getting closer to the optimal manifold. Vice-versa, if the measured force is smaller than the optimal one for the same speed (point B in Fig. \ref{fig:optimal_F_v2}), a lower reel-out speed is commanded (line 7 of Algorithm \ref{fig:optimal_F_v2}), leading to a larger traction force (since the apparent wind experienced by the kite increases) and again makes the operating point converge towards the optimal manifold. Finally, when the measured operating conditions are on the manifold, then the reel-out speed is kept constant (point C in Fig. \ref{fig:optimal_F_v2}).

\section{Simulation results}\label{S:simulation results}

\begin{figure}[htb!]
	\centering
	\includegraphics[width=.7\columnwidth]{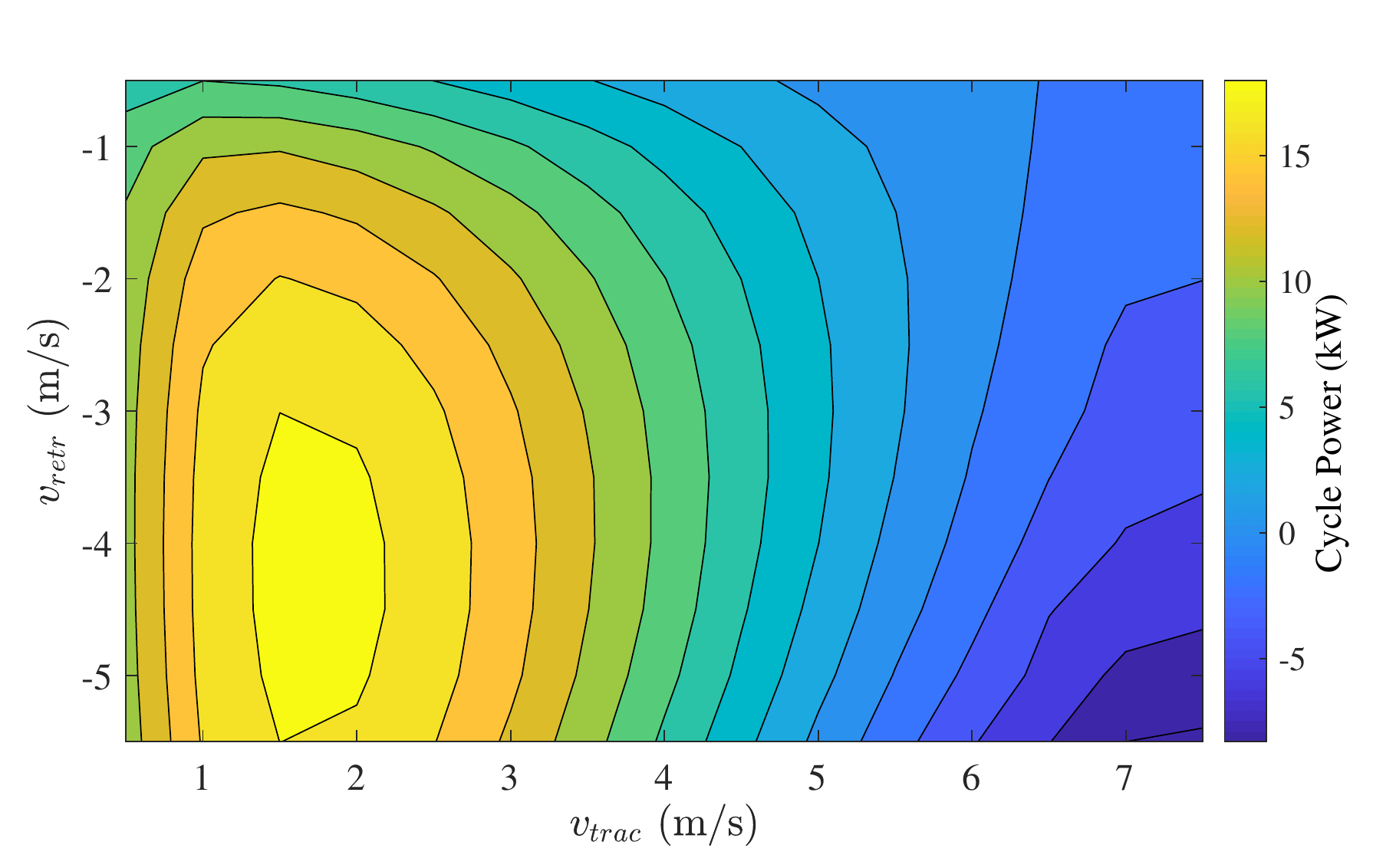}
	\caption{Simulation results. Level curves of the cycle power response surface computed with 9 m/s wind speed.}
	\label{fig:respnse-surface}
\end{figure}

\begin{figure}[htb!]
	\centering
	\includegraphics[width=.7\columnwidth]{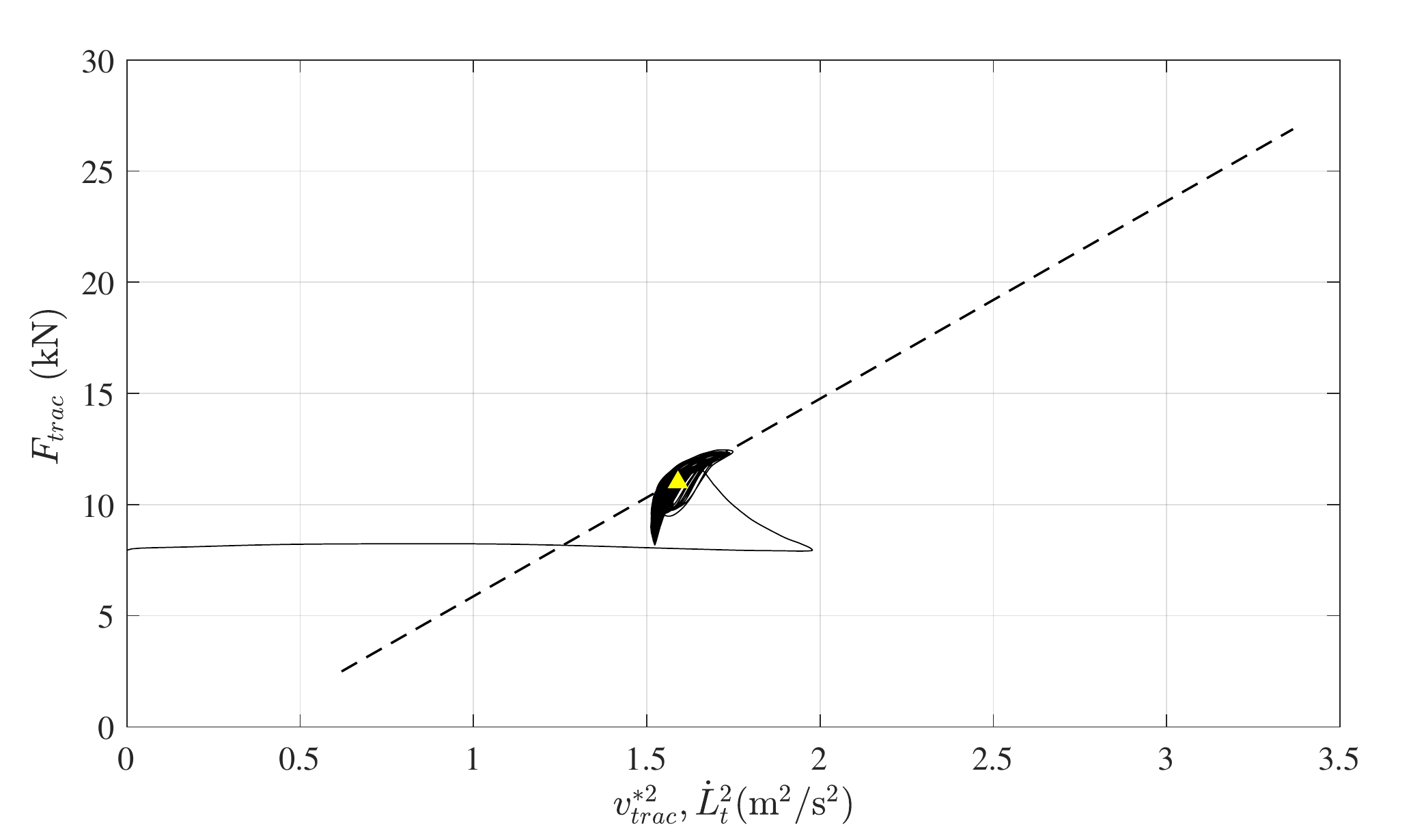}
	\caption{Simulation results. Dashed: optimal manifold of squared tether velocity and tether force. Solid: simulated trajectory of squared tether velocity and force. Yellow triangle: optimal operating condition for the considered wind speed of 6.5 m/s.}
	\label{fig:speed-force-sim}
\end{figure}

\begin{figure}[hbt!]
	\centering 
	\includegraphics[width=.5\textwidth]{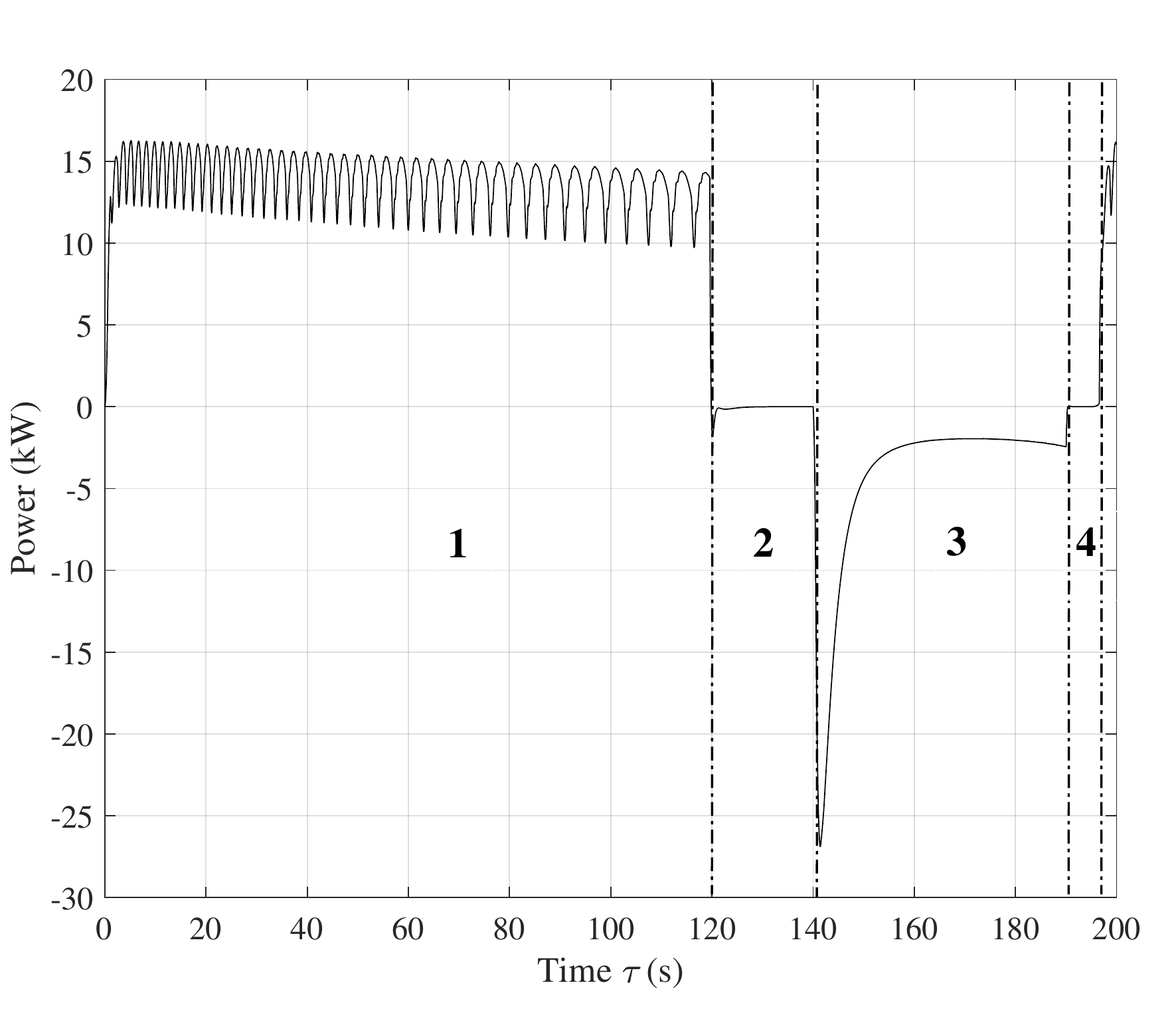}
	\caption{Simulation results. Course of mechanical power during a pumping cycle with $v_{w}=$6.5$\,$m/s,using the proposed reeling approach. 1: Traction phase; 2: Transition 1; 3: Retraction phase; 4: Transition 2.}
	\label{fig:power_profile}
\end{figure}

We applied our method to a system with parameters listed in Table \ref{T:system parameters}.
\begin{table}[htb!]
	\caption{System parameters employed in the numerical simulations.}
	\label{T:system parameters}
	\centering
	\begin{tabular}{|l|r|}
		\hline
		\multicolumn{2}{|c|}{\textbf{Kite}}\\\hline		
		Area	& 25 m$^2$\\ \hline			
		Mass 	&		10.5 kg\\ \hline	
		Wingspan &  	10 m\\ \hline	
		Lift coefficient	&	0.95\\ \hline	
		Lift-to-Drag ratio	&	6\\ \hline	
\multicolumn{2}{|c|}{\textbf{Tether}}\\\hline	
			Diameter	&		6 mm\\ \hline	
			Density		&	975 kg/m3\\ \hline	
			Drag coefficient	&	1\\ \hline	
\multicolumn{2}{|c|}{\textbf{Ground station}}\\\hline				
			Maximum torque		&	1244 Nm\\ \hline	
			Total inertia	&		1.7 kgm$^2$\\ \hline	
			Trasmission ratio	&	26\\ \hline	
			Viscous coefficient	&	0.799\\ \hline	
			Drum radius	&		0.3 m\\ \hline	
	\end{tabular}
	
\end{table}
 Fig. \ref{fig:respnse-surface} presents the response surface obtained with wind speed $v_w=9\,$m/s, computed at step 2)-a) of the approach. Note that, even though the function is not convex everywhere, its level sets are convex for positive cycle power values, thus indicating that the subsequent maximization \eqref{eq:optimization problem} returns the global maximum. We carried out approach with $v_w\in[6.5,\,10]\,$m/s and obtained similar results for all the considered wind speed values
The obtained optimal manifolds in the planes $(v_{trac}^{2},F_{trac})$ and $(v_{trac}^{2},F_{trac})$ are those reported in Fig. \ref{fig:optimal_F_v2}. Fig. \ref{fig:speed-force-sim} presents, as an example, the simulated courses of the actual tether speed and tether force during the traction phase cycle with 6.5 m/s wind speed.
It can be noted that the on-line reeling strategy is able to drive the system's operation on the optimal manifold. Fig. \ref{fig:power_profile}  presents the corresponding course of mechanical power during the whole cycle.\\
In Table \ref{optimal_data_tab} we report a comparison between our method and two other possible alternative approaches: one aiming at maximizing the traction power (\textit{optimal \(P_{trac}\)} approach in Table \ref{optimal_data_tab}), as done e.g. in \cite{Erhard2015,Zgraggen2016}, and the other aiming at reducing the retraction phase as much as possible 
(\textit{fast retraction} approach in Table \ref{optimal_data_tab}). 
The latter approach could be a sensible strategy as it aims to improve the pumping duty cycle. As expected, the results in the Table confirm that the presented approach outperforms the other two candidates, achieving in some cases 25\%-30\% larger cycle power.
Finally, Table \ref{Algo_noAlgo_power} presents a comparison between our approach, and an ideal approach where the wind speed is known, and the optimal reeling speeds are selected accordingly: it can be noted that the proposed method achieves almost optimal performance, however without wind speed measurement.\\


\begin{table}[htb!]
	\caption{Comparison among the proposed approach (optimal $P_{cycle}$), traction power maximization (optimal $P_{trac}$), and fastest retraction phase (fast retraction).}
	\label{optimal_data_tab}
\centering
\begin{tabular}{|c|c|c|c|c|}
\hline
\(v_{w}\) (m/s) & \(v_{trac}\) (m/s) & \(v_{retr}\) (m/s) & P (kW) & Approach \\ \hline
\multirow{3}{*}{6.5} & 1.27 & -3.18 &\textbf{7.6} & optimal \(P_{cycle}\) \\
 & 2.07 & -3.18 & 6.0 & optimal \(P_{trac}\)\\
 & 1.27 & -5 & 2.7 & fast recovery\\ \hline
\multirow{3}{*}{7} & 1.39 & -3.43 & \textbf{9.2} & optimal \(P_{cycle}\)\\
 & 2.23 & -3.43 & 7.5 & optimal \(P_{trac}\)\\
 & 1.39 & -5 & 4.9 & fast recovery\\ \hline
\multirow{3}{*}{7.5} & 1.40 & -3.55 &\textbf{ 11.5} & optimal \(P_{cycle}\)\\
 & 2.39 & -3.55 & 9.2 & optimal \(P_{trac}\)\\
 & 1.40 & -5.00 & 7.8 & fast recovery\\ \hline
\multirow{3}{*}{8} & 1.48 & -3.93 & \textbf{13.5} & optimal \(P_{cycle}\)\\
 & 2.55 & -3.93 & 11.0 & optimal \(P_{trac}\)\\
 & 1.48 & -5.00 & 7.8 & fast recovery\\ \hline
\multirow{3}{*}{8.5} & 1.54 & -4.03 & \textbf{16.22} & optimal \(P_{cycle}\)\\
 & 2.71 & -4.03 & 13.16 & optimal \(P_{trac}\)\\
 & 1.54 & -5.5 & 14.97 & fast recovery\\ \hline
\multirow{3}{*}{9} & 1.68 & -4.34 & \textbf{19.02} & optimal \(P_{cycle}\)\\
 & 2.87 & -4.34 & 15.48 & optimal \(P_{trac}\)\\
 & 1.68 & -5.5 & 18.09 & fast recovery\\ \hline
\multirow{3}{*}{9.5} & 1.77 & -4.47 & \textbf{22.12} & optimal \(P_{cycle}\)\\
 & 3.02 & -4.47 & 18.11 & optimal \(P_{trac}\)\\
 & 1.77 & -6 & 20.77 & fast recovery\\ \hline
\multirow{3}{*}{10} & 1.83 & -4.71 & \textbf{25.56} & optimal \(P_{cycle}\)\\
 & 3.18 & -4.71 & 20.93 & optimal \(P_{trac}\)\\
 & 1.83 & -6.00 & 24.63 & fast recovery\\ \hline
\end{tabular}

\end{table}

\begin{table}[hbt!]
\centering
\begin{tabular}{|c|c|c|}
\hline
\multirow{2}{*}{\(v_{w}\) {[}m/s{]}} & \multicolumn{2}{c|}{\(P_{cycle}\) [kW]} \\ \cline{2-3} 
 & Proposed approach & Optimal value\\ \hline
6.5 & 7.3 & 7.6 \\ \hline
7 & 9.0 & 9.2 \\ \hline
7.5 & 11.1 & 11.5 \\ \hline
8 & 13.4 & 13.5 \\ \hline
8.5 & 16.0 & 16.2 \\ \hline
9 & 18.8 & 19.0 \\ \hline
9.5 & 22.0 & 22.1 \\ \hline
10 & 25.5 & 25.6 \\ \hline
\end{tabular}
\caption{Comparison between the cycle power obtained using the proposed approach and the optimal one assuming perfect knowledge of the wind speed.}
\label{Algo_noAlgo_power}
\end{table}


\section{Conclusions and future developments}

A new design methodology for the tether reeling controller of pumping Airborne Wind Energy systems using soft kites has been presented. The approach is based on the computation of an optimal force-squared speed manifold using a model of the system, and on a feedback control strategy that exploits such a manifold to compute the reference reeling speed on the basis of the measured tether force. Simulation results with a well-established model show that the proposed approach achieves optimal cycle power performance without the need to measure or estimate the wind speed. Future developments are concerned with the study of a similar approach for AWE systems with rigid wing \cite{RSOH19,Todeschini_submitted} and with the experimental test of the method.
\clearpage
\bibliographystyle{plain} 
\bibliography{bibliography}

\end{document}